\documentclass{cs19proc}
\usepackage{txfonts}

\editors{G.~A. Feiden}
\publisher{Zenodo}
\conference{The 19th Cambridge Workshop on Cool Stars, Stellar Systems, and the Sun}
\conferencedate{2016}

\title{Spectroscopic characterisation of CARMENES target candidates from FEROS, CAFE and HRS high-resolution spectra}
\author{V. M. Passegger$^{1}$, 
        A. Reiners$^{1}$,
        S. V. Jeffers$^{1}$, 
        S. Wende$^{1}$,
        P. Sch\"{o}fer$^{1}$,
        P. J. Amado$^{2}$,
        J. A. Caballero$^{3}$,
        D. Montes$^{4}$,
        R. Mundt$^{5}$,
        I. Ribas$^{6}$,
        A. Quirrenbach$^{3}$,
        and the CARMENES Consortium}

\affiliation{$^{1}$ Institut f\"{u}r Astrophysik, Georg-August-Universit\"{a}t G\"{o}ttingen, Friedrich-Hund-Platz 1, D-37077 G\"{o}ttingen, Germany,  \\
		$^{2}$ Instituto de Astrof\'{\i}sica de Andaluc\'{\i}a (CSIC), Glorieta de la Astronom\'{\i}a s/n, E-18008 Granada, Spain, \\
		$^{3}$ Landessternwarte, Zentrum f\"{u}r Astronomie der Universit\"{a}t Heidelberg, K\"{o}nigstuhl 12, D-69117 Heidelberg, Germany, \\
		$^{4}$ Departamento de Astrof\'{\i}sica, Facultad de F\'{\i}sica, Universidad Complutense de Madrid, E-28040 Madrid, Spain, \\
		$^{5}$ Max-Planck-Institut f\"{u}r Astronomie, K\"{o}nigstuhl 17, D-69117 Heidelberg, Germany, \\
		$^{6}$ Institut de Ci\`encies de l'Espai (IEEC-CSIC), Campus UAB, C/Can Magrans s/n, E-08193 Bellaterra, Barcelona, Spain}

\shorttitle{Spectroscopic characterisation of CARMENES target candidates}
\shortauthors{V.M. Passegger et al.}

\abs{CARMENES (\textit{Calar Alto high-Resolution search for M dwarfs with Exoearths with Near-infrared and optical \'{E}chelle Spectrographs}) started a new 
planet survey on M-dwarfs in January this year. The new high-resolution spectrographs are operating in the visible and near-infrared at Calar Alto 
Observatory. They will perform high-accuracy radial-velocity measurements (goal 1 m\,s$^{-1}$ ) of about 300 M-dwarfs with the aim to detect low-mass 
planets within habitable zones.
We characterised the candidate sample for CARMENES and provide fundamental parameters for these stars in order to constrain planetary properties 
and understand star-planet systems. Using state-of-the-art model atmospheres (PHOENIX-ACES) and $\chi2$-minimization with a downhill-simplex method we 
determine effective temperature, surface gravity and metallicity [Fe/H] for high-resolution spectra of around 480 stars of spectral types M0.0-6.5V 
taken with FEROS, CAFE and HRS. We find good agreement between the models and our observed high-resolution spectra.
We show the performance of the algorithm, as well as results, parameter and spectral type distributions for the CARMENES candidate sample, which 
is used to define the CARMENES target sample. We also present first preliminary results obtained from CARMENES spectra. }
\begin{document}

\maketitle

\section{Introduction}
 The new CARMENES instrument is mounted at the 3.5 m telescope at Calar Alto Observatory, located in the Sierra de los Filabres in southern Spain. It consists 
 of two fibre-fed high-resolution spectrographs, operating in the visible wavelength range from 0.52 to 0.96 $\mu$m and in the near-infrared from 0.96 to 1.71 $\mu$m, 
 having a spectral resolution of R~>~80,000. \citep{Quirrenbach2010,Quirrenbach2012,Quirrenbach2014} Both spectrographs will simultaneously perform high-accuracy radial-velocity 
 measurements of about 300 M dwarfs during three years of guaranteed observing time. The aim is to detect low-mass planets within the habitable zones of these stars. \\

For science preparation over 1500 high-resolution spectra have been observed with FEROS, CAFE and HRS to determine effective temperature, surface gravity and metallicity. 
These parameters are fundamental for characterising star-planet systems. The spectra of M dwarfs are very complex, with molecular lines forming due to the low temperatures. 
This makes it difficult to use a line-by-line approach and requires a full spectral synthesis, which in turn necessitates for accurate models that take into account 
the formation of molecules. We use the latest generation PHOENIX model grid, the PHOENIX ACES models \citep{Husser2013}. These models are especially designed for low 
temperature stellar atmospheres and use a new equation of state to accurately reproduce molecular lines. 

\section{Methods and Data}

\begin{center}
  \begin{table*}
    \caption{Properties of spectrographs and data analysed}
    \label{tab:obs}
    \centering %
    \begin{tabular}{cccccc}
      \hline \hline 
      Name & Resolution & Coverage [nm] & No. Spectra & No. Stars & Observing Period\\ 
	   \hline 
      CAFE  & \textasciitilde 65,000 & 396-950  & 623 & 236 &  2013-01-21 to 2014-09-26\\
      FEROS & 48,000 & 350-920 & 455 & 217 & 2012-12-31 to 2014-07-11\\
      HRS   & 60,000 & 420-1100 & 93 & 29 & 2011-09-29 to 2013-06-18\\

      \hline
    \end{tabular}
  \end{table*}
\end{center}

Table~\ref{tab:obs} summarizes the properties of the spectrographs used for observation and the data taken. 
Some observed spectra could not be used for analysis because of different issues, e.g. very low signal-to-noise, observation of wrong target, polluting light from close companions. 

\begin{figure*}
	\centering
	\includegraphics[width=0.85\linewidth]{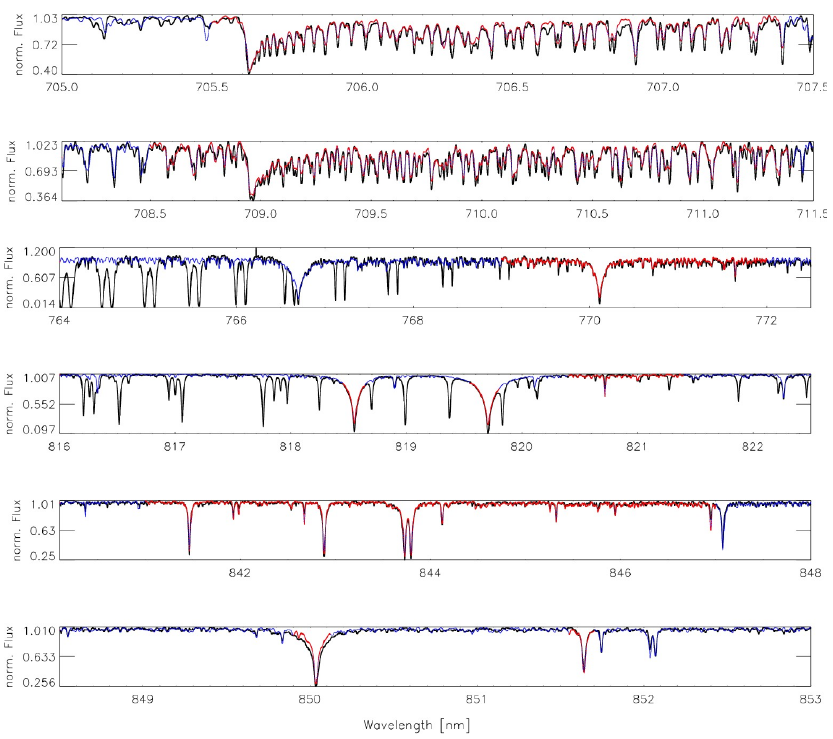}
	\caption{Spectrum of BD+44 2051 (M1.5V, black) and the best fit model (blue: model outside fit region, red: model inside fit regions for $\chi2$-minimization.}
	\label{fig:fit}
\end{figure*}

The method we use was described in detail in \cite{Passegger2016a}. We fit PHOENIX ACES model spectra to our observed spectra. This is done for different spectral 
ranges, including the $\gamma$- and $\epsilon$-TiO  bands (sensitive to temperature and metallicity), the K- and Na-doublets around 768 nm and 819 nm (sensitive to surface gravity and 
metallicity) and two CaII-lines. Rotational velocities determined by \cite{Jeffers2016} are included to account for line broadening due to stellar rotation. Other 
than \cite{Passegger2016a} a downhill simplex is implemented for linear interpolation between the model grid points and a $\chi2$ -minimization determines the best fit to the data. 
Figure~\ref{fig:fit} shows an example fit to CARMENES data.

\section{Results and Discussion}

\begin{figure*}
	\centering
	\includegraphics[width=0.85\linewidth]{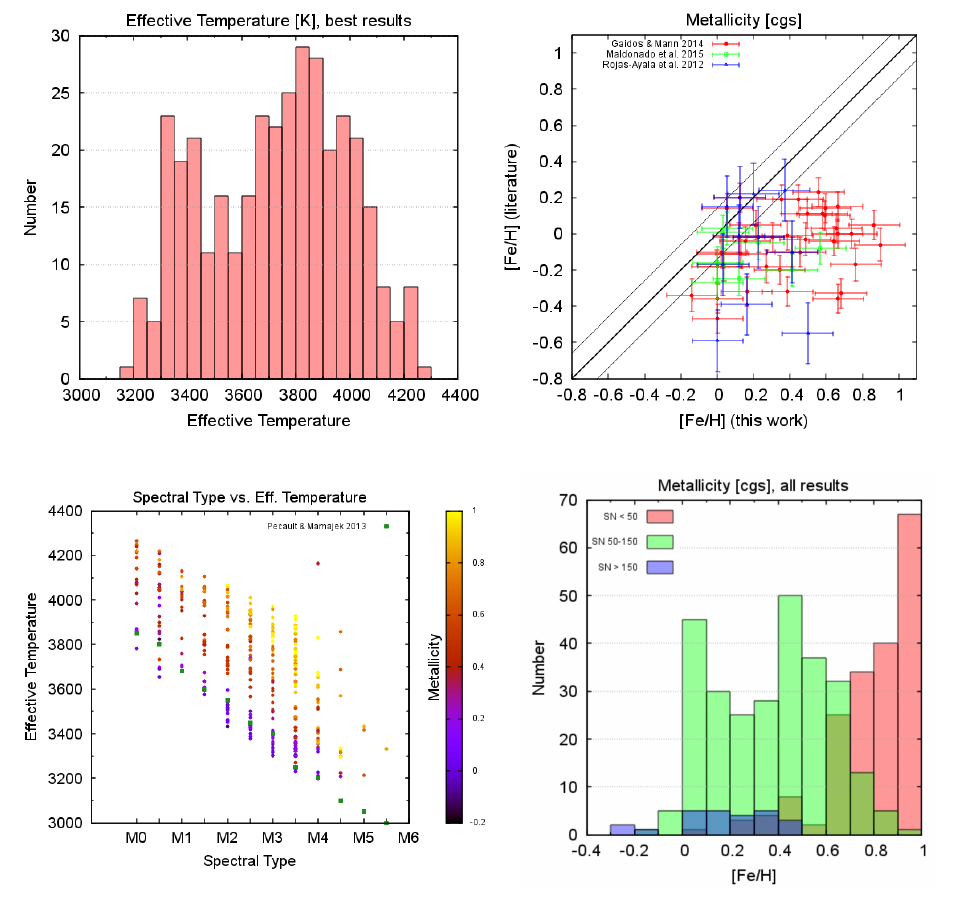}
	\caption{Temperature distribution of candidate sample (upper left), literature comparison for metallicity (upper right), spectral type-temperature relation 
	(lower left, green dots: literature values for solar metallicity found by \cite{PecautMamajek2013}), metallicity distribution for stars observed with FEROS, 
	CAFE and HRS for different SNRs (lower right).}
	\label{fig:results}
\end{figure*}

We obtained stellar parameters for 351 stars from 977 spectra. We find that most stars lie within 3200-3900 K, corresponding to spectral types M1V-M5V, as shown in the upper 
left panel of Figure~\ref{fig:results}. The higher the metallicity the higher the temperature for each spectral type (Figure~\ref{fig:results}, lower left panel). This is consistent with results by 
\cite{Mann2015}. They showed that with increasing metallicity the radius increases, for fixed temperature. The spectral types have been calculated using spectral indices 
\cite{Schoefer2015}. The green squares correspond to a literature computation by \cite{PecautMamajek2013} for solar metallicity. 
A literature comparison with \cite{RojasAyala2012}, \cite{GaidosMann2014} and \cite{Maldonado2015} shows that our values for metallicity turn out to be higher than 
published ones. (Figure~\ref{fig:results}, upper right). One possible explanation for this is that PHOENIX ACES models still cannot reproduce the full depths of some lines (see Figure~\ref{fig:fit}, 4th 
wavelength range), 
which might cause the algorithm to choose higher metallicity models to fit the lines. On the other hand it seems that the signal-to-noise ratio is also very important for 
parameter determination. 75 percent of the stars with [Fe/H] higher than 0.6 have SNRs lower than 50. We find good agreement with expected [Fe/H] values for SNR>50 (Figure~\ref{fig:results}, lower right). 
For the first four months of CARMENES data we find that the parameters show better agreement with literature, having better SNRs.

\section*{Acknowledgments}
{CARMENES is an instrument for the Centro Astron\'{o}mico Hispano-Alem\'{a}n de Calar Alto (CAHA). CARMENES was funded by the German Max-Planck-Gesellschaft (MPG),
the Spanish Consejo Superior de Investigaciones Cient\'{\i}cas (CSIC), the European Union through European Regional Fund (FEDER/ERF), Spanish Ministry 
of Economy and Competitiveness, the state of Baden-W\"{u}rttemberg, the German Science Foundation (DFG), the Junta de Andaluc\'{\i}a, and by the Klaus Tschira
Stiftung, with additional contributions by the members of the CARMENES Consortium (Max-Planck-Institut f\"{u}r Astronomie, Instituto de Astrof\'{\i}sica de Andaluc\'{\i}a, 
Landessternwarte K\"{o}nigstuhl, Institut de Ci\`{e}ncies de l'Espai, Institut f\"{u}r Astrophysik G\"{o}ttingen, Universidad Complutense de Madrid, Th\"{u}ringer Landessternwarte 
Tautenburg, Instituto de Astrof\'{\i}sica de Canarias, Hamburger Sternwarte, Centro de Astrobiolog\'{\i}a, and the Centro Astron\'{o}mico Hispano-Alem\'{a}n).}

\bibliographystyle{cs19proc}
\bibliography{passegger.bib}

\end{document}